\documentclass[amsmath,amssymb,11pt,superscriptaddress,reprint, preprintnumbers, notitlepage,aps,twocolumn,nofootinbib]{revtex4-1}
\pdfoutput=1 
\usepackage[utf8]{inputenc}
\usepackage[english]{babel}
\usepackage{amsmath}
\usepackage{graphicx}
\usepackage{dcolumn}
\usepackage{pbox}
\usepackage{amssymb}
\usepackage{epsfig}
\usepackage{slashed}
\usepackage{amssymb}
\usepackage{ mathrsfs }
\usepackage{color}
\usepackage[font=small]{caption}
\usepackage[font=small]{subcaption}
\usepackage{url}

\definecolor{MyLightBlue}{rgb}{0.22,0.51,0.9}
\definecolor{BrickRed}{rgb}{0.8, 0.25, 0.33}

\RequirePackage{hyperref}
\hypersetup{colorlinks, citecolor=blue,linkcolor=BrickRed, urlcolor=MyLightBlue}

\newcommand\scalemath[2]{\scalebox{#1}{\mbox{\ensuremath{\displaystyle #2}}}}

\makeatletter
\renewcommand\@makecaption[2]{%
  \par
  \vskip\abovecaptionskip
  \begingroup
  
   \small\rmfamily
    \begingroup
     \samepage
     \flushing
     \let\footnote\@footnotemark@gobble
     \@make@capt@title{#1}{#2}\par
    \endgroup
  \endgroup
  \vskip\belowcaptionskip
}
\makeatother

\DeclareUnicodeCharacter{2212}{-}
\begin{document}

\title{\Large Implications of the Zero 1-3 Flavour Mixing  \\ Hypothesis: Predictions for $\theta_{23}^\mathrm{PMNS}$ and  $\delta^\mathrm{PMNS}$
}

\author{\bf Stefan Antusch}
\email[E-mail: ]{stefan.antusch@unibas.ch}
\author{\bf Kevin Hinze}
\email[E-mail: ]{kevin.hinze@unibas.ch}
\author{\bf Shaikh Saad}
\email[E-mail: ]{shaikh.saad@unibas.ch}

\affiliation{Department of Physics, University of Basel, Klingelbergstrasse\ 82, CH-4056 Basel, Switzerland}

\begin{abstract}
\vspace{1mm}
We revisit mixing sum rule relations in the lepton and quark sectors under the assumption that the 1-3 elements of the flavour mixing matrices ($V^u_L,V^d_L,V^e_L,V^\nu_L$) are zero in the flavour basis. We consider the exact relations resulting from the validity of this ``zero 1-3 flavour mixing hypothesis'' and analyse their implications based on the current experimental data, including effects from RG running. In particular, we analyse how the existing precise measurement of $\theta_{13}^\mathrm{PMNS}$ allows to derive predictions for $\theta_{23}^\mathrm{PMNS}$ in models with constrained $\theta_{12}^\mathrm{e}$. As examples, we calculate the predictions for $\theta_{23}^\mathrm{PMNS}$ which arise in classes of Pati-Salam models and SU(5) GUTs that relate $\theta_{12}^\mathrm{e}$ to $\theta_{12}^\mathrm{d}$. We also derive a novel  ``lepton phase sum rule'', valid under the additional assumption of small charged lepton mixing contributions. We furthermore point out that, in the context of GUT flavour models, the quark and lepton CP violating phases $\delta^\mathrm{CKM}$ and $\delta^\mathrm{PMNS}$ can both be predicted from a single imaginary element in the mass matrices.
\end{abstract}

\maketitle
%%%%%%%%%%%%%%%%%%%%%%%%%%%%%%%%%%%%%%%%%%%%%%%
%%%%%%%%%%%%%%%%%%%%%%%%%%%%%%%%%%%%%%%%%%%%%%%
\section{Introduction}
Mixing sum rule relations are a useful tool for understanding how classes of flavour models can give rise to predictions for the observable parameters of the PMNS and CKM mixing matrices in the lepton and quark sectors. 
By providing such understanding, they can be a valuable guidance for model building. 
In the lepton sector, it has been shown that under the assumption that the charged lepton left mixing angles $\theta_{ij}^{e}$ are small and $\theta_{13}^{e},\theta_{13}^{\nu}\approx 0$, the approximate lepton mixing relations~\cite{King:2005bj,Masina:2005hf,Antusch:2005kw,Antusch:2007rk,Antusch:2012fb}
\begin{align}
& \theta_{12}^{\mathrm{PMNS}} -  \theta_{13}^{\mathrm{PMNS}} \cot{\theta_{23}^{\mathrm{PMNS}}} \cos{\delta^{\mathrm{PMNS}}} \approx  \theta_{12}^{\nu} \:,\\
& \theta_{13}^{\mathrm{PMNS}} \approx \sin(\theta_{23}^{\mathrm{PMNS}}) \theta_{12}^{e}  \label{eq:t13t23}
\end{align}
can be derived. The relations show that flavour models which satisfy the assumptions and predict a value of $\theta_{12}^{\nu}$ or $\theta_{12}^{e}$, inevitably predict a correlation among the parameters of the PMNS matrix, i.e.\ here among the mixing angles $\theta_{ij}^{\mathrm{PMNS}}$ and the Dirac CP phase $\delta^{\mathrm{PMNS}}$. Crucially, apart from the one model dependent quantity ($\theta_{12}^{\nu}$ or $\theta_{12}^{e}$), the relations involve only quantities measurable in neutrino oscillation experiments.

In the quark sector, analogous relations, derived under the condition of hierarchical up and down quark mass matrices (i.e.\ small left mixing angles $\theta_{ij}^{u}$ and $\theta_{ij}^{d}$) and $\theta_{13}^{u},\theta_{13}^{d}\approx 0$ read \cite{Antusch:2009hq}:
\begin{align}
& \left| \theta^\textrm{CKM}_{12}-\frac{\theta^\textrm{CKM}_{13}}{\theta^\textrm{CKM}_{23}}e^{-i\delta^\textrm{CKM}} \right| \approx \theta^d_{12}  \:, \\
& \frac{\theta_{13}^{\mathrm{CKM}}}{\theta_{23}^{\mathrm{CKM}} } \approx \theta_{12}^{u} \:.
\end{align}
Furthermore, a ``quark phase sum rule'' has been derived, which relates the two model-dependent phases of the 1-2 mixings in the up- and down-quark sectors to the observable CP violation via the relation \cite{Antusch:2009hq}
\begin{align}
\delta_{12}^d-\delta^u_{12}  \approx \alpha\approx \frac{\pi}{2}\:,  \label{quarkPSR0}
\end{align}
 where $\alpha$ is the unitarity triangle angle, with current measured value $\alpha= \left(91.98^{+0.82}_{-1.40} \right)^\circ$ at $2\sigma$ CL \cite{CKMFIT}, defined as
\begin{align}
\alpha &= \arg\left(-\frac{V_{td}V_{tb}^\ast}{V_{ud}V_{ub}^\ast}  \right) \\
&\approx \arg\left(1-\frac{\theta^{\mathrm{CKM}}_{12}\theta^{\mathrm{CKM}}_{23}}{\theta^{\mathrm{CKM}}_{13}}e^{-i\delta^{\mathrm{CKM}}}  \right). \label{quarkPSR}
\end{align}

The goal of this paper is to revisit these sum rule relations and to analyse their predictive power in the light of the current experimental data. We focus on the following aspects:
\begin{itemize}
    \item To start with, we note that a minimal assumption, under which relations can be derived that do not depend on unphysical phases, is that of zero 1-3 elements of the flavour mixing matrices in the flavour basis, or, in terms of the most common parametrization, that the individual 1-3 flavour mixing angles $\theta_{13}^i$, with $i\in\{u,d,e,\nu\}$ vanish (approximately). We will apply this zero 1-3 flavour mixing hypothesis to obtain exact versions of the mixing sum rule relations (cf.\ also \cite{Ballett:2014dua,Marzocca:2013cr,Petcov:2014laa}). 
    \item We then address the question how one can use these relations, given the improved experimental data on the leptonic mixing parameters, to understand how predictions for the PMNS parameters can arise in classes of flavour models that satisfy the zero 1-3 flavour mixing hypothesis. In particular, we will show that the existing  precise measurement of $\theta_{13}^{\mathrm{PMNS}}$ leads to predictions for $\theta_{23}^{\mathrm{PMNS}}$ (which currently has a rather wide experimentally allowed range), from relations between $\theta_{12}^{e}$ and $\theta_{12}^{d}$ that can emerge in Pati-Salam models and SU(5) GUTs.   
    We also discuss the predictions for $\delta^{\mathrm{PMNS}}$ in models that feature a fixed value of  $\theta_{12}^{\nu}$.
    \item Furthermore, extending the concept of phase sum rules from the quark sector to the lepton sector, we derive a novel  ``lepton phase sum rule''. We find, however, that in contrast to the mixing sum rule relations, the phase sum rules in lepton (quark) sectors only hold free of unphysical phases in a small angle approximation for the charged lepton (quark) mixing contributions. 
    \item Finally, applying the phase sum rules in both sectors, we point out that in the context of GUT flavour models, the quark and lepton CP violating phases can both be predicted from a single imaginary element in the mass matrices. 
\end{itemize}

The paper is organized as follows: In Sec.~\ref{sec2}, we fix the notation for the parametrizations of unitary matrices used in this work. In Sec.~\ref{sec3}, we motivate and discuss the zero 1-3 flavour mixing hypothesis. Exact mixing sum rules in the lepton and quark sectors are discussed in Secs.~\ref{sec4} and~\ref{sec5}, respectively. Predictions for $\theta_{23}^\mathrm{PMNS}$ in quark-lepton unified models are exemplified in Sec.~\ref{sec6}. Furthermore, we derive a phase sum rule in the lepton sector in Sec.~\ref{sec7} and examine the possible unified origin of the CP-violating Dirac phases in the quark and leptons sectors. Finally, we conclude in Sec.~\ref{sec9}.

%%%%%%%%%%%%%%%%%%%%%%%%%%%%%%%%%%%%%%%%%%%%%%%
\section{Parametrization of PMNS and CKM matrices}\label{sec2}

%%%%%%%%%%%%%%%%%%%%%%%%%%%%%%%%%%%%%%%%%%%%%%%
\textbf{Parametrization of PMNS matrix:}--
The part of the Lagrangian that contains lepton masses is given by
\begin{align}
-\mathcal{L}_Y\supset \overline \ell_L M_e \ell_R + \frac{1}{2} \overline{\nu^c_L}M_\nu \nu_L + h.c.,    
\end{align}
which are diagonalized as follows:
\begin{align}
&M_e= V^e_L \textrm{ diag($m_e, m_\mu, m_\tau$) } V^{e \dagger}_R, \label{Me}\\
&M_\nu= V^{\nu \ast}_L \textrm{ diag($m_{\nu_1}, m_{\nu_2}, m_{\nu_3}$) } V^{\nu \dagger}_L.
\end{align}
Then the mixing matrix in the lepton sector $U$ arises in the leptonic charged current interactions in the mass basis,
\begin{align}
&-\mathcal{L}_{cc}=\frac{g}{\sqrt{2}} \overline \ell_L \gamma^\mu U \nu_L W^-_\mu +h.c.,   
\end{align}
where, 
\begin{align}
U= V^{e \dagger}_L V^{\nu}_L.    \label{U}
\end{align}
The leptonic mixing matrix $U$ can be written as,
\begin{align}
&U=P \; U_\textrm{PMNS},
\end{align}
where $P$ is a diagonal phase matrix containing three unphysical phases $P=\textrm{diag}\{e^{i\phi_1}, e^{i\phi_2}, e^{i\phi_3}\}$ and $U_\textrm{PMNS}$ is the 
Pontecorvo–Maki–Nakagawa–Sakata (PMNS) matrix for which we use the PDG parametrization \cite{ParticleDataGroup:2020ssz}:
\begin{align}
&U_\textrm{PMNS}=R_{23}U_{13}R_{12}Q, 
\\
&R_{23}=\begin{pmatrix}
1&0&0\\
0&c_{23}&s_{23}\\
0&-s_{23}&c_{23}
\end{pmatrix},\;
U_{13}= \scalemath{0.9}{
\begin{pmatrix}
c_{13}&0&s_{13}e^{-i\delta_{CP}}\\
0&1&0\\
-s_{13}e^{i\delta_{CP}}&0&c_{13}
\end{pmatrix} },
\\
&
R_{12}=\begin{pmatrix}
c_{12}&s_{12}&0\\
-s_{12}&c_{12}&0\\
0&0&1
\end{pmatrix},\;
Q=\begin{pmatrix}
e^{i\beta_1}&0&0\\
0&e^{i\beta_2}&0\\
0&0&1
\end{pmatrix},
\end{align}
where we have used $c_{ij}=\cos \theta_{ij}^\textrm{PMNS}$, $s_{ij}=\sin \theta_{ij}^\textrm{PMNS}$. $U_\textrm{PMNS}$ contains three physical phases: one Dirac CP-violating phase $\delta^\textrm{PMNS}$ and two Majorana phases $\beta_{1,2}$ (these latter two phases become unphysical if neutrinos are Dirac particles). In this work, we assume neutrinos to be Majorana in nature.

%%%%%%%%%%%%%%%%%%%%%%%%%%%%%%%%%%%%%%%%%%%%%%%
\vspace{0.5cm}
\textbf{Parametrization of CKM matrix:}--
Similarly, the part of the Lagrangian that contains quark masses is given by
\begin{align}
-\mathcal{L}_Y\supset \overline u_L M_u u_R + \overline d_L M_d d_R + h.c.,    
\end{align}
which are diagonalized by,
\begin{align}
&M_u= V^u_L \textrm{ diag($m_u, m_c, m_t$) } V^{u \dagger}_R, \label{Me}\\
&M_d= V^d_L \textrm{ diag($m_d, m_s, m_b$) } V^{d \dagger}_R.
\end{align}
The corresponding mixing matrix in the quark sector $V$ arises from charged quark current interactions in the mass basis,
\begin{align}
&-\mathcal{L}_{cc}=\frac{g}{\sqrt{2}} \overline u_L \gamma^\mu V d_L W^+_\mu +h.c.,   
\end{align}
where 
\begin{align}
V= V^{u \dagger}_L V^d_L.    \label{V}
\end{align}
The quark mixing matrix $V$ can be parametrised as 
\begin{align}
&V=P^\prime \; V_\textrm{CKM} Q^\prime,
\\
&V_\textrm{CKM}=R^\textrm{CKM}_{23}U^\textrm{CKM}_{13}R^\textrm{CKM}_{12},\label{CKM}
\end{align}
where $V_\textrm{CKM}$ is called the 
Cabibbo-Kobayashi-Maskawa (CKM) mixing matrix that contains a single (Dirac) physical phase in $U^\textrm{CKM}_{13}$, which we denote by $\delta^\textrm{CKM}$. Then two diagonal phase matrices $P^\prime$ and $Q^\prime$ contain five unphysical phases.

%%%%%%%%%%%%%%%%%%%%%%%%%%%%%%%%%%%%%%%%%%%%%%%
\vspace{0.5cm}
\textbf{Alternative parametrization of unitary matrices:}--  It is often convenient to use an alternative parametrization for the unitary matrices that diagonalize the mass matrices of the individual flavour sectors, making use of the fact that a general $3\times 3$ unitary matrix $W$, containing nine parameters (three angles and six phases), can be parametrized as 
\begin{align}
&W=W_{23}W_{13}W_{12}S, \label{W1}
\end{align}
where three unitary matrices $W_{ij}$ are defined as
\begin{align}
W_{23}= \begin{pmatrix}
1&0&0\\
0&c_{23}&s_{23}e^{-i\delta_{23}}\\
0&-s_{23}e^{i\delta_{23}}&c_{23}
\end{pmatrix},
\end{align}
and analogously for the other two matrices, and where $S$ is a diagonal phase matrix $S=\textrm{diag}\{e^{i\chi_1}, e^{i\chi_2},e^{i\chi_3}\}$. 
Throughout this work, we use this general parametrization Eq.~\eqref{W1} for the two unitary matrices $V^e_L, V^\nu_L$  in the leptonic sector (Eq.~\eqref{U}) as well as for $V^u_L, V^d_L$ in the quark sector (Eq.~\eqref{V}).\footnote{For a similar parametrization, cf.\ Ref.~\cite{King:2002nf}.}

%%%%%%%%%%%%%%%%%%%%%%%%%%%%%%%%%%%%%%%%%%%%%%%
\section{The Zero 1-3 Flavour Mixing Hypothesis}\label{sec3}
The fact that in both sectors, quarks and leptons, the 1-3 elements of the CKM and PMNS mixing matrices are much smaller than the other elements motivates the hypothesis that the 1-3 elements of all flavour mixing matrices $V^u_L,V^d_L,V^e_L,V^\nu_L$ in the up-quark, down-quark, charged lepton and neutrino sectors, or, in terms of the most common parametrization, that the individual 1-3 flavour mixing angles, vanish (approximately). With left mixing angles $\theta^{u,d,e,\nu}_{13}=0$ in the flavour basis (equivalent to $W^{u,d,e,\nu}_{13}=\mathcal{I}_{3\times 3}$), the 1-3 mixings in the observable CKM and PMNS mixing matrices are then induced by the non-commutativity of the individual flavour rotations. In the following, we will revisit the mixing sum rule relations under this assumption.

We note that although we will work here with the exact hypothesis, our results remain valid as approximate statements when the $\theta^{x}_{13}$ are sufficiently small. One example is the case when $\theta^{u,d,e}_{13} \approx 0$ is realised with hierarchical mass matrices via ``texture zeros'' in the 1-3 positions of the mass matrices (with LR convention as used in this paper). In the neutrino sector, $\theta^{\nu}_{13} \approx 0$ can e.g.\ be realised in the framework of ``Sequential Dominance (SD)'' \cite{King:1999mb,King:2002nf,Antusch:2004gf,Antusch:2010tf} for neutrino masses via the type I seesaw mechanism, when the dominant right-handed neutrino coupling vector to lepton doublets has a zero in the $e$-position and is orthogonal to the subdominant one, in the limit of the third right-handed neutrino being approximately decoupled. This is the case e.g.\ in Constrained Sequential Dominance \cite{King:2005bj}. 
In the following, we will study the implications of the zero 1-3 flavour mixing hypothesis.

%%%%%%%%%%%%%%%%%%%%%%%%%%%%%%%%%%%%%%%%%%%%%%%
\begin{figure*}[t!]
\centering
\includegraphics[width=0.3\textwidth]{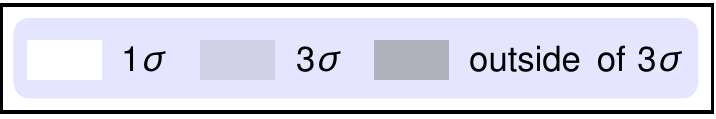}\\\vspace{0.5cm}

\includegraphics[width=0.43\textwidth]{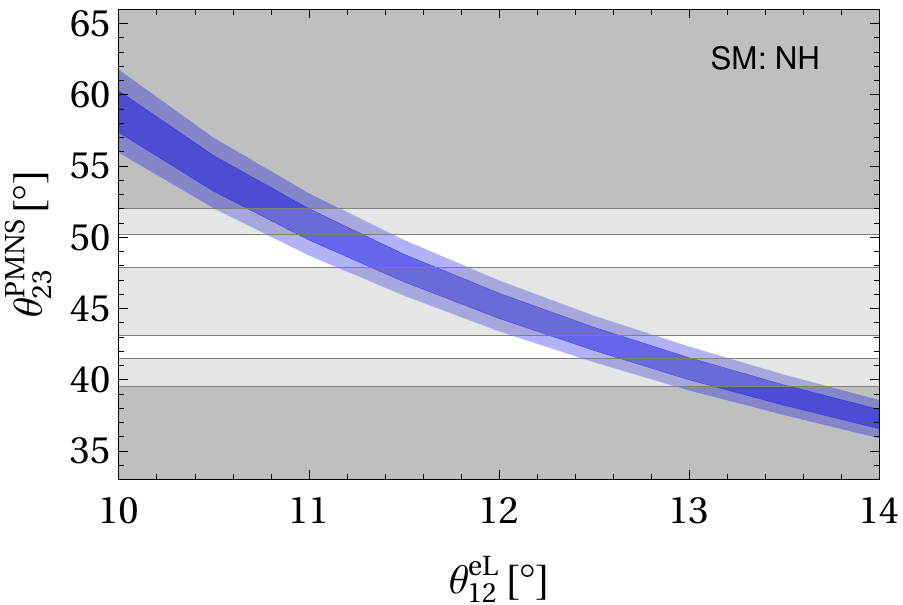}
\hspace{0.5cm}
\includegraphics[width=0.43\textwidth]{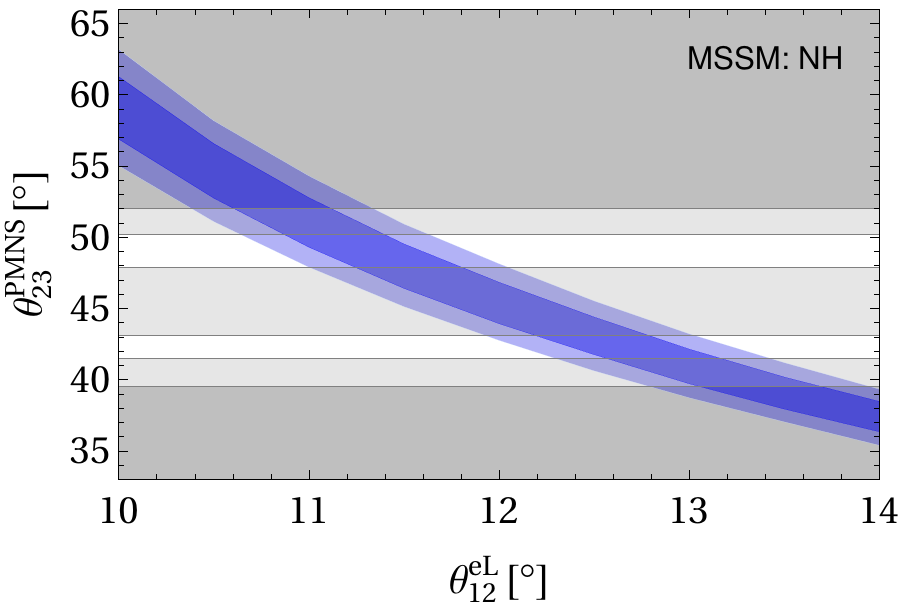}\vspace{0.5cm}
\includegraphics[width=0.43\textwidth]{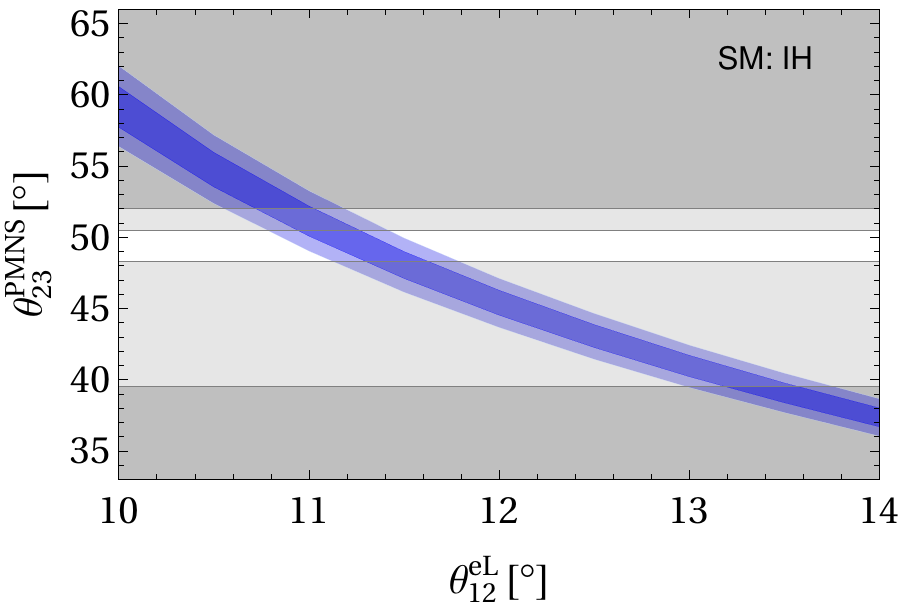}
\hspace{0.5cm}
\includegraphics[width=0.43\textwidth]{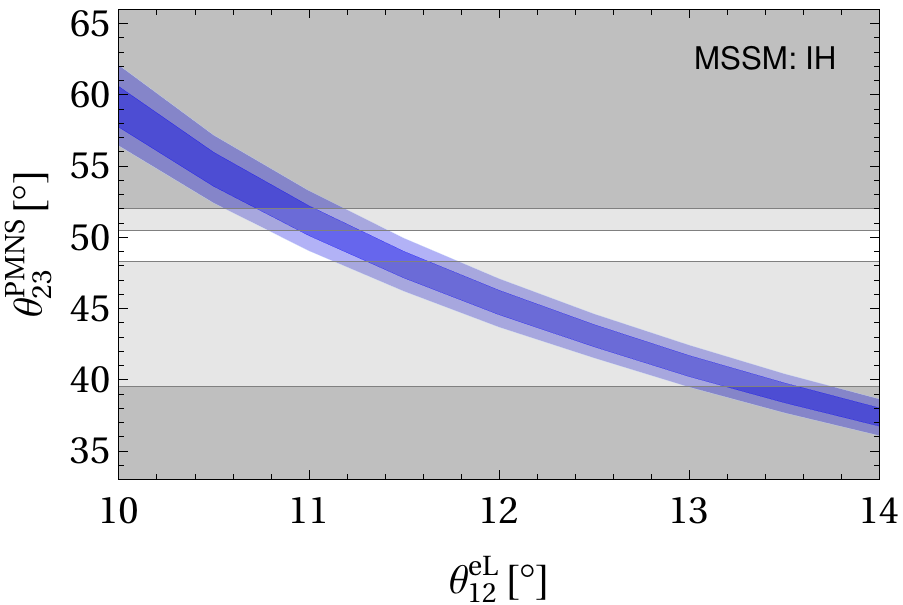}
\caption{Predictions for the 1$\sigma$ and 2$\sigma$ HPD intervals of the low scale PMNS mixing angle $\theta_{23}^\textrm{PMNS}$ as a function of the GUT scale parameter $\theta_{12}^e$, including RG effects. The white region indicates the current experimental 1$\sigma$ range, whereas the light gray region represents the current experimental 3$\sigma$ range. Left: SM. Right: MSSM. Top: NH. Bottom: IH.} \label{fig:t23}
\end{figure*} 

%%%%%%%%%%%%%%%%%%%%%%%%%%%%%%%%%%%%%%%%%%%%%%%
\section{Exact Mixing Sum Rule Relations in the Lepton Sector}\label{sec4}
%%%%%%%%%%%%%%%%%%%%%%%%%%%%%%%%%%%%%%%%%%%%%%%
\textbf{Exact relation between $\theta_{13}^{\mathrm{PMNS}}$, $\theta_{23}^{\mathrm{PMNS}}$ and $\theta^e_{12}$:}-- 
In this section, we consider an exact form of the relation between $\theta_{23}^{\mathrm{PMNS}}, \theta_{13}^{\mathrm{PMNS}}$ and $\theta_{12}^{e}$  \cite{King:2005bj,Masina:2005hf,Antusch:2005kw,Antusch:2007rk,Antusch:2012fb}, which arises under the zero 1-3 flavour mixing hypothesis when we divide in Eq.~\eqref{U} the (13)-entry by the (23)-entry, leading to (cf.\ also \cite{Ballett:2014dua})
\begin{align}
e^{i(-\delta+\phi_1-\phi_2)}\frac{t_{13}}{s_{23}}
=
-t^e_{12} e^{-i(\delta^e_{12}+\chi^e_1-\chi^e_2)}.\label{sum1}
\end{align}
Here we have dropped the PMNS labels for brevity, and abbreviated $t_{ij} := \tan (\theta_{ij})$.
By taking the modulus of the above equation we obtain the relation
\begin{align}
s^{\mathrm{PMNS}}_{23} = \frac{t^{\mathrm{PMNS}}_{13}}{ t^e_{12}}.\label{lepton-sum-rule-1}
\end{align}
Currently, $\theta^\textrm{PMNS}_{13}$ is measured with great accuracy, its $1\sigma$ allowed range corresponds to $\left(8.57^{+0.13}_{-0.12}\right)^\circ$ \cite{NUFIT,Esteban:2020cvm}. On the contrary, $\theta^\textrm{PMNS}_{23}$ has large uncertainty associated with its measurements. While in the past the relation of Eq.~(\ref{eq:t13t23}) has been used to derive the value of $\theta^\textrm{PMNS}_{13}$ generated by the charged lepton 1-2 mixing contribution $\theta_{12}^{e}$ (often assuming maximal $\theta_{23}^{\mathrm{PMNS}}$), we point out that it is now more valuable to use it with the precise measurement of $\theta_{13}^{\mathrm{PMNS}}$ to predict the less well measured PMNS mixing angle $\theta_{23}^{\mathrm{PMNS}}$.

\begin{figure}[t!]
    \centering
    \includegraphics[width=0.45\textwidth]{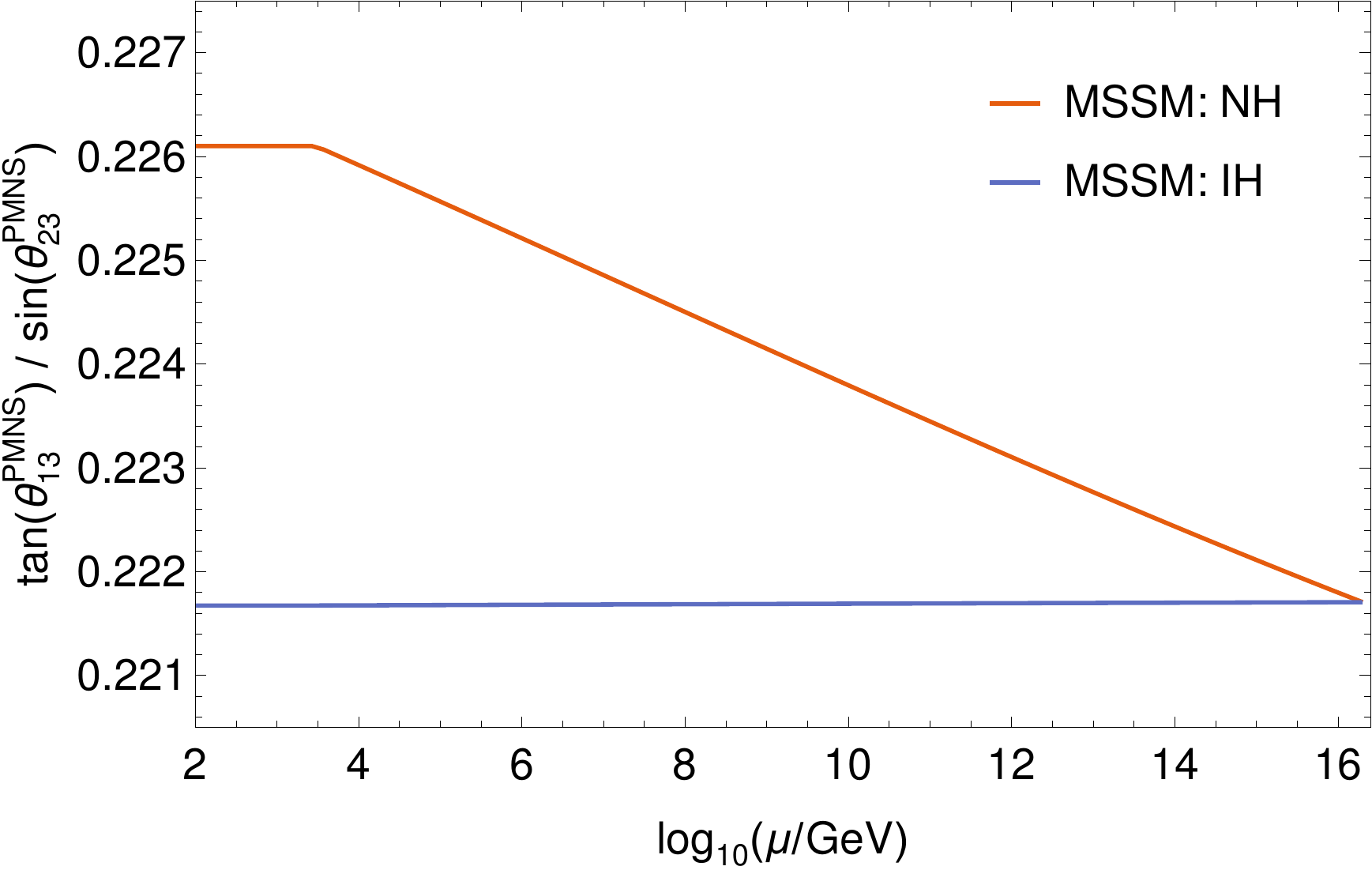}
    \caption{Running of the quotient ${\tan(\theta_{13}^{\mathrm{PMNS}})}/{\sin(\theta_{23}^{\mathrm{PMNS}})}$ for both NH and IH in the MSSM with $\tan \beta = 50$. As it can be seen, for the case of IH the RG effects for $\theta_{13}^{\mathrm{PMNS}}$ and $\theta_{23}^{\mathrm{PMNS}}$ almost cancel each other such that the quotient remains nearly invariant under RG running.}
    \label{fig:insignificant}
\end{figure}

\textbf{Numerical analysis:}--
We can now quantitatively analyse, using the precisely measured value of $\theta^\textrm{PMNS}_{13}$ (cf.\ \cite{NUFIT,Esteban:2020cvm}), how  Eq.~\eqref{lepton-sum-rule-1} leads to a prediction for $\theta^\textrm{PMNS}_{23}$ in terms of $\theta_{12}^{e}$.  However, since the relation Eq.~\eqref{lepton-sum-rule-1} is valid at the flavour scale, which we here set equal to the GUT scale, RG running effects (cf.\ \cite{Antusch:2003kp,Antusch:2005gp} and references therein) must be properly taken into account to find the accurate prediction for $\theta^\textrm{PMNS}_{23}$ at the low scale.\footnote{We note that we assume for our RG analysis a realisation of the neutrino mass matrix by a type I seesaw mechanism where, however, the neutrino Yukawa couplings are significantly smaller than unity such that they do not notably affect the RG evolution. This essentially leaves us with the running of the dimension 5 operator.} 
For this purpose, we have implemented our setup in \texttt{REAP} \cite{Antusch:2005gp} with zero 1-3 mixings for the charged fermion Yukawa coupling matrices $Y^{u,d,e}$ as well as for the neutrino Majorana mass matrix $M_\nu$. For a fixed value of $\theta^e_{12}$ within the range $(10-14)^\circ$, all other parameters (at the high scale) both in the quark and the lepton sectors are freely varied. After running down to the low scale, all observables are then fitted to their experimental values~\cite{Antusch:2013jca,NUFIT}, except $\theta^\textrm{PMNS}_{23}$ for which we obtain a theory prediction. 
We have performed fits and Markov Chain Monte Carlo (MCMC) analyses both in the SM as well as in the MSSM framework for both normal (NH) and inverted (IH) neutrino mass hierarchies, assuming  strong hierarchies with the lightest neutrino mass set to zero. We consider the running from $M_{\rm{GUT}}=2\times 10^{16}$ GeV to $M_{\rm{Z}}=91.18$ GeV. For the case of the MSSM, in our numerical analysis, the SUSY scale is chosen to be 3 TeV, and $\tan\beta$ is taken as a free parameter (which we vary from 10 to 50). Outcomes (i.e.\ predictions for $\theta^\textrm{PMNS}_{23}$) of our fitting procedure as a function of $\theta^e_{12}$ are presented in Fig.~\ref{fig:t23}. 
One can use the plots to read off the predicted range for $\theta^\textrm{PMNS}_{23}$ for models that feature a fixed value of  $\theta^e_{12}$.
As can be seen from these plots, due to $\tan \beta$ effects in the MSSM that can enhance the running, the allowed range of $\theta^e_{12}$ consistent with experimentally measured values of $\theta^\textrm{PMNS}_{23}$ is slightly wider in comparison to the SM scenario, in the case of NH. 
In the case of IH, the two plots for the SM and MSSM look identical. The reason for this is that for IH the running of the ratio ${\tan(\theta_{13}^{\mathrm{PMNS}})}/{\sin(\theta_{23}^{\mathrm{PMNS}})}$ is practically zero, and thus both plots look as if we had just evaluated the formula of Eq.~(\ref{lepton-sum-rule-1}) at low scale. This stability against RG running is illustrated in Fig.~\ref{fig:insignificant} (cf.\ also \cite{Antusch:2012fb}).

Furthermore, we can also state the currently preferred values of $\theta^e_{12}$, given the experimentally allowed region for $\theta^\textrm{PMNS}_{23}$. To this end we have varied the fitting procedure described above by including $\theta^e_{12}$ as a fit parameter and by adding $\theta^\textrm{PMNS}_{23}$ to the list of observables to be fitted. 
Performing MCMC analyses we now obtain the highest posterior density (HPD) intervals for $\theta^e_{12}$ consistent with the experimental data for $\theta^\textrm{PMNS}_{23}$, which are listed in Table~\ref{tab:01}.  These allowed ranges for $\theta^e_{12}$ can be used as a guideline for building flavour models consistent with the current results from $\theta^\textrm{PMNS}_{23}$ measurements.

\begin{table}[h]
    \centering 
    \begin{tabular}{|c|c|c|}\hline
         & $\theta_{12}^e$ (1$\sigma$) & $\theta_{12}^e$ (2$\sigma$) \\\hline
        SM: NH & $10.99^\circ - 11.68^\circ$ & $10.84^\circ - 13.08^\circ$  \\
        & $12.25^\circ - 12.86^\circ$ & \\\hline
        SM: IH & $10.90^\circ - 11.80^\circ$ & $10.81^\circ - 13.02^\circ$  \\\hline
        MSSM: NH & $10.98^\circ - 11.86^\circ$ & $10.75^\circ - 13.18^\circ$  \\
        & $12.17^\circ - 12.72^\circ$ & \\\hline
        MSSM: IH & $10.88^\circ - 11.81^\circ$ & $10.81^\circ - 12.99^\circ$  \\\hline
    \end{tabular}
    \caption{Viable 1$\sigma$ and 2$\sigma$ HPD intervals for $\theta^e_{12}$ which predict the correct experimental value for $\theta^\textrm{PMNS}_{23}$. The RG running effects have been considered in the case of the SM as well as of the MSSM for both NH as well as IH, as discussed in the main text.}
    \label{tab:01}
\end{table}

\vspace{0.5cm}
\textbf{Exact lepton mixing sum rule involving $\theta_{12}^{\nu}$ and $\delta_\textrm{PMNS}$:}--
A second consequence of the zero 1-3 mixing hypothesis is that the value of $\theta^\nu_{12}$, which is a model parameter, can be related to the leptonic Dirac CP phase $\delta_\textrm{PMNS}$ \cite{King:2005bj,Masina:2005hf,Antusch:2005kw,Antusch:2007rk,Antusch:2012fb}. In the following we consider the exact relation which arises when we divide in Eq.~\eqref{U} the (31)-entry by the (32)-entry. This yields (cf.\ also \cite{Ballett:2014dua})
\begin{align}
&
e^{i(\beta_1-\beta_2)}
\frac{-t_{12}t_{23}+e^{i\delta}s_{13}}{t_{23}+e^{i\delta}t_{12}s_{13}}
=
-t^\nu_{12} e^{i(\delta^\nu_{12}+\chi^\nu_1-\chi^\nu_2)},\label{sum2}
\end{align}
where we again dropped the PMNS labels for brevity. Taking the modulus leads to 
\begin{align}
&
t^\nu_{12}=\left|
\frac{t^\textrm{PMNS}_{12}t^\textrm{PMNS}_{23}-e^{i\delta^\textrm{PMNS}}s^\textrm{PMNS}_{13}}{t^\textrm{PMNS}_{23}+e^{i\delta^\textrm{PMNS}}t^\textrm{PMNS}_{12}s^\textrm{PMNS}_{13}} \right|. \label{lepton-sum-rule-2}
\end{align}
Using the experimentally measured values of neutrino mixing angles as well as the Dirac CP phase, one can calculate the preferred value of $\theta^\nu_{12}$, which subsequently can be used as a guideline for flavour model building. Alternatively, since the neutrino mixing angles are much more precisely measured than the Dirac CP phase $\delta^\textrm{PMNS}$, each flavour model that features a fixed value of $\theta^\nu_{12}$ and that satisfies the zero 1-3 mixing hypothesis predicts (utilizing the experimental values of the neutrino mixing angles) a specific Dirac CP phase by which it can be experimentally tested.

\begin{figure*}
    \centering
    \includegraphics[width=0.3\textwidth]{FIG/legend.pdf}\\\vspace{0.5cm}

    \includegraphics[width=0.45\textwidth]{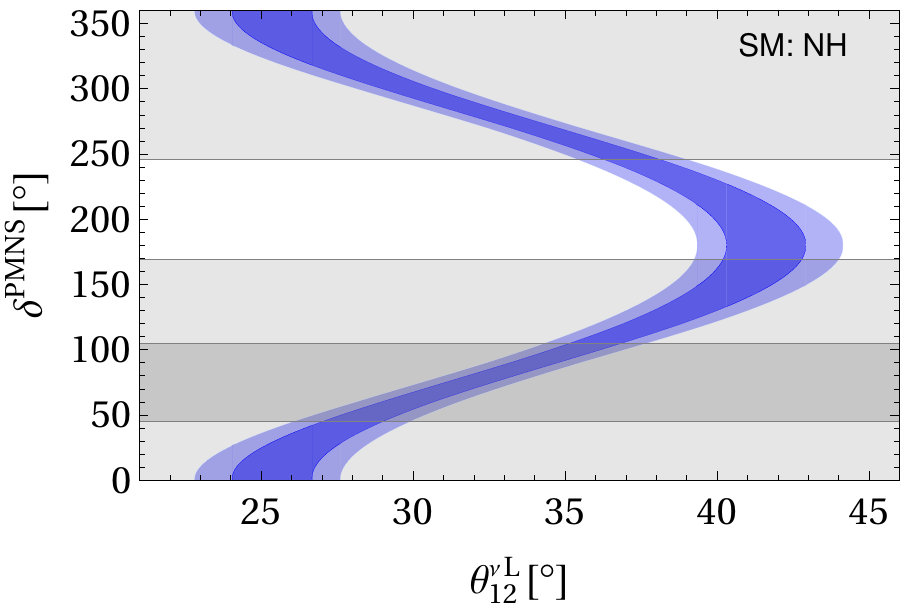}\hspace{5mm}
     \includegraphics[width=0.45\textwidth]{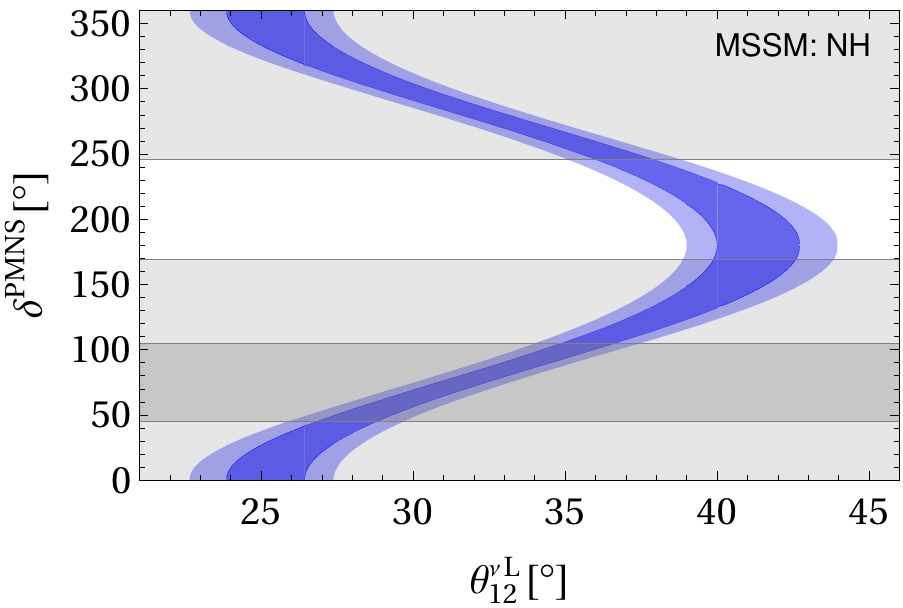}\vspace{5mm}
      \includegraphics[width=0.45\textwidth]{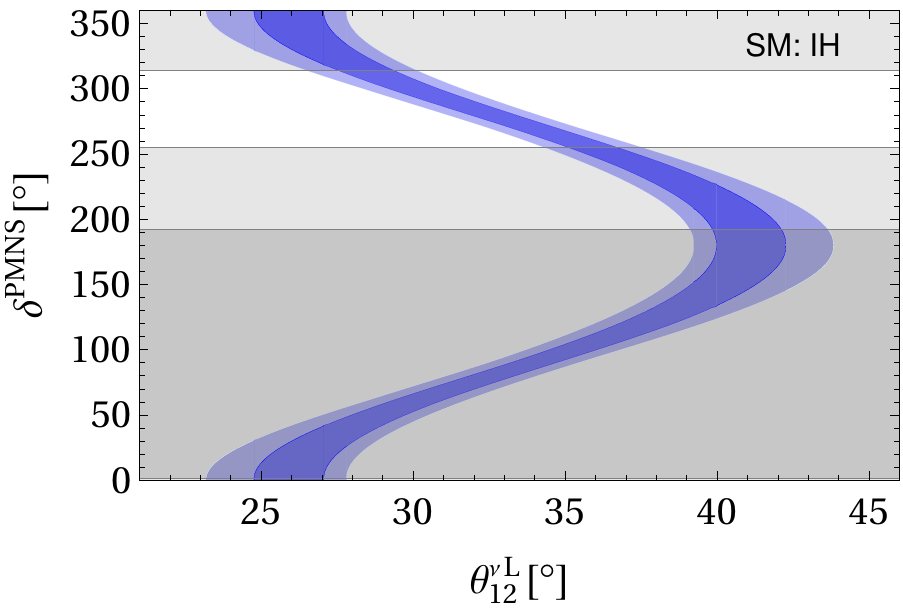}\hspace{5mm}
       \includegraphics[width=0.45\textwidth]{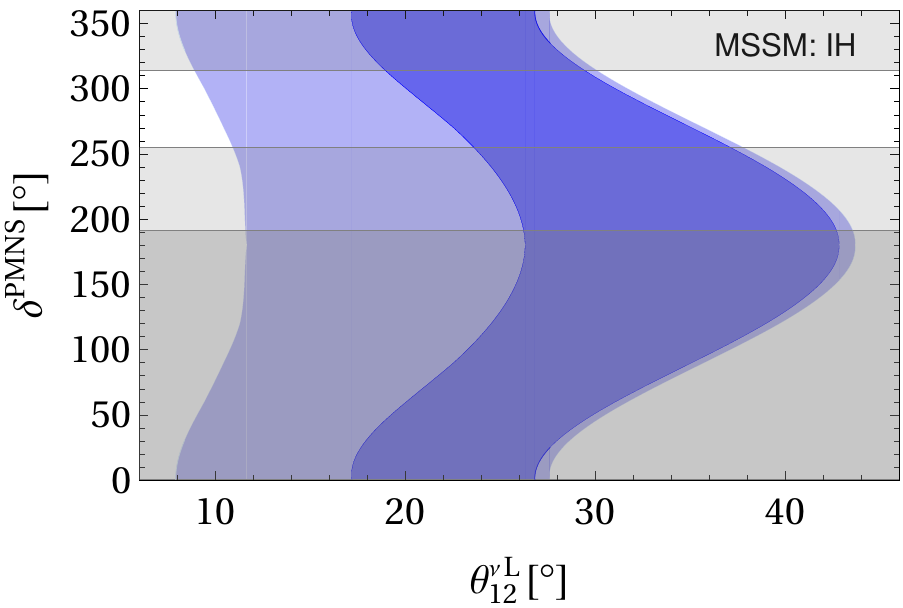}
    \caption{Viable 1$\sigma$ and 2$\sigma$ HPD intervals for $\theta_{12}^{\nu}$ wich predict the correct experimental value for the PMNS Dirac CP phase $\delta^\textrm{PMNS}$ as a function of $\theta_{12}^\nu$ including running effects. The experimental 1$\sigma$ and 3$\sigma$ ranges are represented by white and light gray regions, respectively. Left: SM. Right: MSSM. Top: NH. Bottom: IH.}
    \label{fig:delta-withrunning}
\end{figure*}

\begin{figure}
    \centering
    \includegraphics[width=0.3\textwidth]{FIG/legend.pdf}\\\vspace{0.5cm}
    \includegraphics[width=0.45\textwidth]{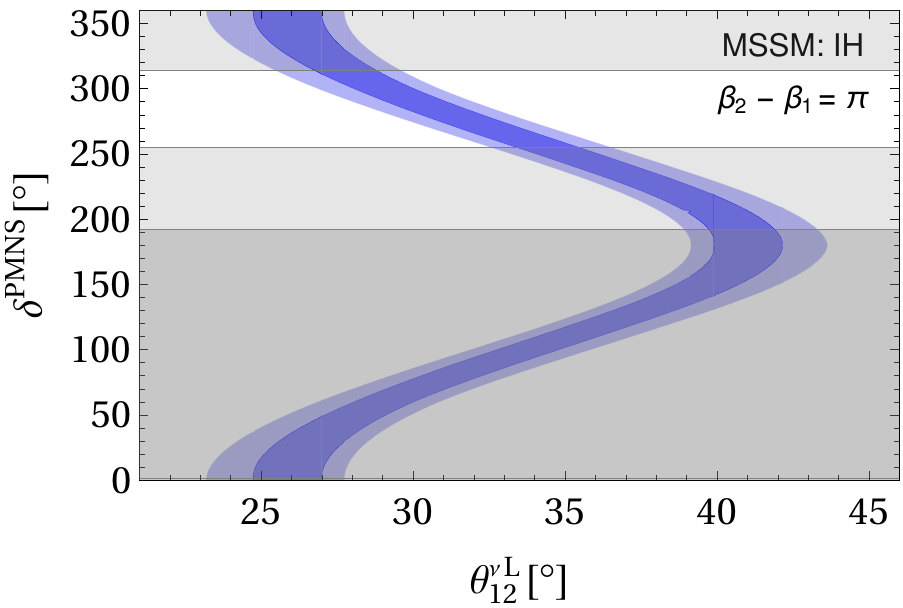}
    \caption{Viable 1$\sigma$ and 2$\sigma$ HPD intervals for $\theta_{12}^{\nu}$ which predict the correct experimental value for the PMNS Dirac CP phase $\delta^\textrm{PMNS}$ as a function of $\theta_{12}^\nu$ including running effects for the MSSM with IH and a Majorana phase difference $\beta_2 - \beta_1 = \pi$. The experimental 1$\sigma$ and 3$\sigma$ ranges are represented by white and light gray regions, respectively.}
    \label{fig:delta-withrunning-MajPhasediff-pi}
\end{figure}

\textbf{Numerical analysis:}-- 
As discussed above, since the sum rule relations such as Eq.\ \eqref{lepton-sum-rule-2} hold at the flavour scale (here taken as the GUT scale), RG effects have to be considered in order to give the low scale prediction for $\delta^\textrm{PMNS}$ in terms of the (high scale) value of $\theta^\nu_{12}$. To obtain the predicted range for $\delta^\textrm{PMNS}$ in terms of $\theta^\nu_{12}$, we have again implemented the zero 1-3 flavour mixing setup in \texttt{REAP}. Keeping $\theta_{12}^\nu$ fixed, we now vary the remaining GUT scale input parameters, compute the RG evolution from the GUT scale $M_{\rm{GUT}}=2\times10^{16}$\ GeV down to the $Z$ scale $M_Z=91.18$\ GeV at which we fit, apart from $\delta^\textrm{PMNS}$, all the observables to the experimental data. 
We have performed fits and MCMC analyses with NH and IH for both the SM and the MSSM. In the latter case, we have again chosen a SUSY scale of 3\ TeV and included $\tan\beta$ in the parameter list (varied from 10 to 50). Fig.\ \ref{fig:delta-withrunning} visualizes our numerical results. The results can be used to read off the predicted range for $\delta^\textrm{PMNS}$ for models with fixed $\theta^\nu_{12}$. One can see that the allowed parameter region is largely extended in the MSSM for IH. This is due to RG running effects which can particularly strongly affect $\theta   ^\textrm{PMNS}_{12}$ when the Majorana phase difference $\beta_2 - \beta_1$ is zero and when $\tan \beta$ is large, as can be seen e.g.\ from the analytical formulae for the running derived in \cite{Antusch:2003kp}.  
On the other hand, in many flavour models with strong inverse neutrino mass hierarchy the Majorana phases satisfy $\beta_2 - \beta_1 \approx \pi$, since this can explain the almost degenerate neutrino masses $m_1$ and $m_2$ from a pseudo-Dirac structure of the neutrino mass matrix. We therefore show in 
Fig.~\ref{fig:delta-withrunning-MajPhasediff-pi} another plot with the results for the MSSM with IH and the Majorana phase difference fixed to $\beta_2 - \beta_1 = \pi$, which can be applied to this class of models. 

Moreover, in order to obtain values of $\theta_{12}^\nu$ preferred by the current experimental data, we have performed additional fits and MCMC analyses for which we added $\theta_{12}^\nu$ to the list of input parameters and also included  $\delta^\textrm{PMNS}$ in the fit to the experimental data. These results are presented in Table~\ref{tab:02} and can provide useful guidance for building models which agree with the current preferred values of the Dirac CP phase $\delta^\textrm{PMNS}$.
Note that the preferred value for $\theta_{12}^\nu$ is different for NH compared to IH also in the SM. The reason for this is the different experimentally preferred range for $\delta^\textrm{PMNS}$ and not an effect coming from enhanced RG running with $\beta_2 - \beta_1 = 0$, which only has a large effect in the MSSM when also $\tan \beta$ is large.

\begin{table}[h]
    \centering 
    \begin{tabular}{|c|c|c|}\hline
         & $\theta_{12}^\nu$ (1$\sigma$) & $\theta_{12}^\nu$ (2$\sigma$) \\\hline
        SM: NH & $36.37^\circ - 43.43^\circ$ & $26.75^\circ - 44.23^\circ$  \\\hline
        SM: IH & $28.11^\circ - 35.64^\circ$ & $25.80^\circ - 39.62^\circ$ \\\hline
        MSSM: NH & $36.24^\circ - 43.29^\circ$ & $26.64^\circ - 44.11^\circ$  \\\hline
        MSSM: IH & $19.06^\circ - 34.22^\circ$ & $8.68^\circ - 37.41^\circ$ \\\hline
    \end{tabular}
    \caption{Viable 1$\sigma$ and 2$\sigma$ HPD intervals for $\theta^\nu_{12}$ which predict the correct experimental value for $\delta^\textrm{PMNS}$. We have considered the RG evolution for the SM and the MSSM for both NH as well as for IH), as discussed in the main text. }
    \label{tab:02}
\end{table}

%%%%%%%%%%%%%%%%%%%%%%%%%%%%%%%%%%%%%%%%%%%%%%%
\section{Exact Mixing Sum Rule Relations in the Quark Sector}\label{sec5}
%%%%%%%%%%%%%%%%%%%%%%%%%%%%%%%%%%%%%%%%%%%%%%%
Exact versions of the quark mixing sum rules \cite{Antusch:2009hq} valid under the assumption of zero 1-3 flavour mixing hypothesis can be obtained similarly to the lepton sector. First, we divide the (13)-entry by the (23)-entry in Eq.~\eqref{V} and then take the absolute values, which yields, 
\begin{align}
t^u_{12}=\frac{t^{\mathrm{CKM}}_{13}}{s^{\mathrm{CKM}}_{23}}.\label{quark-sum-rule-1}
\end{align}
Furthermore, we divide the (31)-entry by (32)-entry in Eq.~\eqref{V} and take the absolute value, which provides, 
\begin{align}
&
t^d_{12}=\left|
\frac{t^\textrm{CKM}_{12}t^\textrm{CKM}_{23}-e^{i\delta^\textrm{CKM}}s^\textrm{CKM}_{13}}{t^\textrm{CKM}_{23}+e^{i\delta^\textrm{CKM}}t^\textrm{CKM}_{12}s^\textrm{CKM}_{13}} \right|. \label{quark-sum-rule-2}
\end{align}
Since CKM parameters are very precisely measured in the experiments, these relations essentially fix the 1-2 mixings. The results for both, the SM and MSSM cases, are $\theta_{12}^u\in (4.75^\circ,\,5.14^\circ)$ and $\theta_{12}^d\in (11.85^\circ,\,12.40^\circ)$ at $1\sigma$.

%%%%%%%%%%%%%%%%%%%%%%%%%%%%%%%%%%%%%%%%%%%%%%%
\section{Example Predictions for $\theta_{23}^{\mathrm{PMNS}}$ in Quark-Lepton Unified Models}\label{sec6}
%%%%%%%%%%%%%%%%%%%%%%%%%%%%%%%%%%%%%%%%%%%%%%%

Predictions for $\theta^e_{12}$ can arise e.g.\ from quark-lepton unification in Pati-Salam~\cite{Pati:1973rp,Pati:1974yy} or GUT models~\cite{Georgi:1974sy,Georgi:1974yf, Georgi:1974my,Fritzsch:1974nn}, which can dictate strong correlations between down-type ($Y_d$) and charged-lepton ($Y_e$) Yukawa matrices. To establish such predictive mass relations from quark-lepton unification, we assume that the entries of $Y_d$ and $Y_e$ are each generated by one single joint GUT operator (also referred to as ``single operator dominance'', cf.\ \cite{Antusch:2011qg,Antusch:2012fb,Antusch:2019avd}). Consequently, the matrix entries are linked only via group theoretical Clebsch factors (cf.\ \cite{Antusch:2009gu,Antusch:2013rxa}).

To illustrate how this assumption can lead to predictions for $\theta^e_{12}$, and thus for  $\theta_{23}^{\mathrm{PMNS}}$, we focus on the 1-2 sector of the down-type and charged lepton mass matrices, ignoring possible 2-3 rotations (which are here taken to be small). 
Writing the 1-2 block of the $Y_d$ matrix as
\begin{align}
Y_d=\begin{pmatrix}
y_{11}&y_{12}\\
y_{21}&y_{22}
\end{pmatrix},
\end{align}
$Y_e$ then, at the unification scale, has the form
\begin{align}
Y_e^{SU(5)}=\begin{pmatrix}
c_{11}y_{11}&c_{21}y_{21}\\
c_{12}y_{12}&c_{22}y_{22}
\end{pmatrix},\;\;\;
Y_e^{PS}=\begin{pmatrix}
c_{11}y_{11}&c_{12}y_{12}\\
c_{21}y_{21}&c_{22}y_{22}
\end{pmatrix},
\end{align}
for SU(5) GUT and Pati-Salam (PS) partial unification, respectively. Here, $c_{ij}$ are the Clebsch factors that relate the $Y_d$ and $Y_e$ entries. For example, possible options for the Clebsch factors in SU(5) GUTs are $|c_{ij}|\in \{1/6,\,4/9,\,1/2,\,2/3,\, 1,\, 3/2,\,2,\, 3,\, 9/2,\, 6,\, 9,\, 18\}$ and in PS models $|c_{ij}|\in \{1/3,\,3/4,\, 1,\,3/2,\, 2,\, 3,\, 9\}$ \cite{Antusch:2009gu,Antusch:2013rxa}. These coefficients are constrained by the requirement that the down-quark and charged-lepton masses of the first two generations have to be consistent with the measured values at low energy, taking RG running and possible SUSY threshold corrections into account.
From the above Yukawa matrices, one can infer a relation between the left 1-2 mixing angles $\theta^d_{12}$ (fixed from the zero 1-3 flavour mixing hypothesis to be $\theta_{12}^d\in (11.85^\circ,\,12.40^\circ)$ and $\theta^e_{12}$ (cf.\ also \cite{Antusch:2011qg}). 

For the PS case, in the limit of hierarchical Yukawa matrices with $y_e \ll y_\mu$ and $y_d \ll y_s$, one finds 
\begin{align}
&\tan (\theta^{e, PS}_{12})\approx  \left| \frac{c_{12}y_{12}}{c_{22}y_{22}} \right| \approx \left| \frac{c_{12}}{c_{22}}\right| \tan(\theta^d_{12})\:,   \label{eq-theta12e PS}
\end{align}
which directly implies a prediction for $\theta^{e}_{12}$ in terms of $|{c_{12}}/{c_{22}}|$ and the allowed range for $\theta_{12}^d$. 

In SU(5) GUTs, without any further constraint, the analogous relation would additionally depend on the ratio $|{y_{21}}/{y_{12}}|$, since $Y^{SU(5)}_e$ is related to $Y_d^T$ and not as in PS models directly to $Y_d$. Predictivity can however be restored by, e.g., requiring a zero 1-1 element of $Y_d$ (and thus also of $Y_e$). 
Then, again in the limit of hierarchical Yukawa matrices where $y_\mu \approx c_{22} y_{22}$, $y_e = c_{21} y_{21}c_{12}y_{12}/(c_{22} y_{22})$, $\tan(\theta^d_{12}) = y_{12}/y_{22}$ and $\tan(\theta^{e, SU(5)}_{12})=c_{21}y_{21}/(c_{22}y_{22})$, we can write
\begin{align}
& \frac{y_e}{y_\mu} \approx  \left|\frac{c_{21}y_{21}c_{12}y_{12}}{(c_{22} y_{22})^2}\right| \approx  \left|\frac{c_{12}}{c_{22}}\right| \tan(\theta^{e, SU(5)}_{12}) \tan(\theta^d_{12})\:.
\end{align}
We further note that the Yukawa coupling ratio $y_e/y_\mu$ is very stable under RG running and also under possible SUSY threshold corrections (cf.\ discussion in \cite{Antusch:2013jca}), such that we can evaluate it using the running low scale masses (or, alternatively, the running Yukawa couplings at $M_Z$ provided in \cite{Antusch:2013jca}). This leads to the relation 
\begin{align}\label{tan12d-tan12e}
& \tan(\theta^{e, SU(5)}_{12}) \approx  \left|\frac{c_{22}}{c_{12}}\right| \frac{m_e}{m_\mu} \frac{1}{\tan(\theta^d_{12})} \:.
\end{align}
which implies a prediction for $\theta^{e}_{12}$, depending only on the Clebsch factor ratio $|{c_{22}}/{c_{12}}|$, the precisely known ratio $m_e/m_\mu$, and the allowed range for $\theta_{12}^d$.

Among the Clebsch factors mentioned above and discussed in \cite{Antusch:2009gu,Antusch:2013rxa}), only a few lead (utilizing Eq.\ \eqref{lepton-sum-rule-1}) to predictions for $\theta_{23}^{\mathrm{PMNS}}$ compatible with the current experimental data. Two promising combinations of Clebsches in SU(5) are $c_{22} = 9/2,\,c_{12}=1/2$ and $c_{22} = 6,\,c_{12}=2/3$, which yield the same ratio $|{c_{22}}/{c_{12}}|$ and thus identical predictions for $\theta_{23}^{\mathrm{PMNS}}$, and $c_{22} = 9/2,\,c_{12}=4/9$. We like to remark that the Clebsch factor $4/9$ stems from a higher order operator, as discussed in \cite{Antusch:2013rxa}, whereas the other Clebsch factors appear at dimension 5. Furthermore, in PS, Clebsch combinations with $|c_{12}|=|c_{22}|$ lead to a (currently) viable prediction for $\theta_{23}^{\mathrm{PMNS}}$. 

The three choices of Clebsch factors mentioned above, combined with the resulting values for $\theta_{12}^e$ as well as their respective predicted range for $\theta_{23}^\text{PMNS}$ for the example case of the MSSM with NH, are shown in Table\ \ref{tab-predictions for theta23pmns}.
The predictions for $\theta_{23}^\text{PMNS}$ are obtained from an MCMC analysis considering the RG running effects in the MSSM with NH and utilizing the HPD results for $\theta_{12}^d$.

\begin{table}[ht]
    \centering
    \renewcommand{\arraystretch}{1.5}
    \begin{tabular}{|l|ccc|ccc|}
    \hline
         SU(5) & ($c_{22},c_{12}$) & $\theta_{12}^e$ (1$\sigma$) &&& $\theta_{23}^\text{PMNS}$ (1$\sigma$)&  \\\hline
         & ($\frac{9}{2},\frac{1}{2}$) & $10.97^\circ-11.49^\circ$ &&& $47.04^\circ-51.65^\circ$& \\ 
         & ($\frac{9}{2},\frac{4}{9}$) & $12.31^\circ-12.88^\circ$ &&& $40.67^\circ-44.16^\circ$& 
         \\\hline\hline
         PS & ($c_{22},c_{12}$) & $\theta_{12}^e$ (1$\sigma$) &&& $\theta_{23}^\text{PMNS}$ (1$\sigma$)& \\\hline
         & (3,3) & $11.85^\circ-12.40^\circ$ &&& $43.05^\circ-47.07^\circ$&  \\\hline
    \end{tabular}
    \caption{SU(5) and PS Clebsch factor combinations with predicted ranges of $\theta_{12}^e$, together with their respective predictions for $\theta_{23}^\text{PMNS}$. As mentioned in the main text, the Clebsch combination ($9/2,1/2)$ can be replaced by ($6,2/3$) giving the same prediction for $\theta_{23}^\text{PMNS}$.}
    \label{tab-predictions for theta23pmns}
\end{table}

We like to remark that, at this level of model independence, there are comparatively large uncertainties in the predictions. They are, to some extend, caused by the uncertainty in the amount of RG running, which could be significantly reduced if $\tan \beta$ was known. Additionally, a further reduction of the uncertainties in the predictions for $\theta_{23}^\text{PMNS}$ could be achieved by even more precise measurements of $\theta_{13}^\text{PMNS}$ and of the CKM parameters (reducing the allowed range of $\theta_{12}^d$). 

Furthermore, the predictions are also expected to get more precise when the Clebsch factor combinations are extended to specify the other relevant Yukawa matrix entries. For instance in the SU(5) case with zero 1-1 elements of $Y_{d}$ (and $Y_{e}$), fixing the Clebsch factor $c_{21}$ leads to an additional prediction for the ratio of the running masses $m_s/m_d$ from the relation $m_s/m_d \approx y_{22}^2/(y_{12}y_{21}) \approx (m_\mu/m_e)(c_{12}c_{21}/c_{22}^2)$, which has to be taken into account when fitting a specific scenario, with all Clebsch factors fixed and model-specific restrictions implemented, to the experimental data. 

Such fits have been performed in \cite{Antusch:2018gnu}, and one can see from the tables provided there that the best-fit values for $\theta^{e}_{12}$ are sensitive to the choice of the additional Clebsch factor $c_{21}$, but remain within the ranges given in Table\ \ref{tab-predictions for theta23pmns}. Cases considered in \cite{Antusch:2018gnu} include the possibility to complement the choice $c_{22} = 9/2,\,c_{12}=1/2$ by $c_{21} = 9/2$ or $3$, or $c_{22} = 6,\,c_{12}=2/3$ by $c_{21} = 9/2$ or $6$. We note a possible completion of the Clebsch combination $c_{22} = 9/2,\,c_{12}=4/9$ could be $c_{21} = 9/2$. Finally, we remark that a possible way to judge the viability of Clebsch combinations is provided by the double ratio $d=(y_\mu/y_e)(y_d/y_s)$ \cite{Antusch:2013jca}, which is very stable under RG running also also under possible SUSY threshold corrections. It can here be evaluated as $c_{22}^2/(c_{12}c_{21})$ and is experimentally constrained as $d=10.7^{+1.8}_{-0.8}$ at $1\sigma$.

%%%%%%%%%%%%%%%%%%%%%%%%%%%%%%%%%%%%%%%%%%%%%%%
\section{A Phase Sum Rule in the Lepton Sector}\label{sec7}
%%%%%%%%%%%%%%%%%%%%%%%%%%%%%%%%%%%%%%%%%%%%%%%
\textbf{Leptonic phase sum rule:}--
In this section we derive a phase sum rule in the lepton sector, which has not been discussed in the literature before. In contrast to the mixing sum rules discussed above, we find a relation that does not contain unphysical phases only in the limit of small angle approximation for the charged lepton mixings (i.e.\ requiring small $\theta^e_{ij}$), which is thus an extra assumption. Then, to the leading order, the (11)-, (22)-, and (33)-entries in Eq.~\eqref{U} lead to the following relations,   
\begin{align}
e^{i(\beta_1+\phi_1)}c_{12}c_{13}&\approx e^{i(\chi^\nu_1-\chi^e_1)}c^\nu_{12}+ ...\;, 
\\
e^{i(\beta_2+\phi_2)}c_{12}c_{23}&\approx e^{i(\chi^\nu_2-\chi^e_2)}c^\nu_{12}c^\nu_{23}+ ...\;, 
\\
e^{i\phi_3}c_{13}c_{23}&\approx e^{i(\chi^\nu_3-\chi^e_3)}c^\nu_{23}+ ...\;, 
\end{align}
where the ellipses represent next-to-leading order terms. These relations can be used to relate the unphysical phases, 
\begin{align}
&\phi_1=\chi^\nu_1-\chi^e_1-\beta_1,
\\&
\phi_2=\chi^\nu_2-\chi^e_2-\beta_2,
\\&
\phi_3=\chi^\nu_3-\chi^e_3.
\end{align}
Using these relations and combining Eqs.~\eqref{sum1} and \eqref{sum2}, we obtain (dropping the PMNS labels for brevity) 
\begin{align}
&
e^{i(\delta_{12}^\nu-\delta^e_{12})}t^e_{12}t^\nu_{12}=\frac{t_{13}}{s_{23}}
\frac{s_{13}-t_{12}t_{23}e^{-i\delta}}{t_{23}+t_{12}s_{13}e^{i\delta}},
\end{align}
which yields
\begin{align}
& \delta_{12}^\nu-\delta^e_{12}
=
\arg\left(  
\frac{s^\textrm{PMNS}_{13}-t^\textrm{PMNS}_{12}t^\textrm{PMNS}_{23}e^{-i\delta^\textrm{PMNS}}}{t^\textrm{PMNS}_{23}+t^\textrm{PMNS}_{12}s^\textrm{PMNS}_{13}e^{i\delta^\textrm{PMNS}}}
\right).\label{lepton-phase-sum-rule}
\end{align}
This new  phase sum rule in the lepton sector determines the difference $\delta_{12}^\nu-\delta^e_{12}$ between two model-dependent phases. In models satisfying the zero 1-3 flavour mixing hypothesis, this can provide insight into the possibilities for predicting $\delta^\textrm{PMNS}$ (the least well measured PMNS parameter) in flavour models. Preferred ranges of $\delta_{12}^\nu-\delta^e_{12}$ from MCMC analyses are summarized in Table~\ref{tab:04}. 

\begin{table}[h]
    \centering 
    \begin{tabular}{|c|c|c|}\hline
         & $\delta_{12}^{\nu}-\delta_{12}^e$ (1$\sigma$) & $\delta_{12}^{\nu}-\delta_{12}^e$ (2$\sigma$) \\\hline
        SM: NH & $297.9^\circ - 14.6^\circ$ & $217.3^\circ - 37.6^\circ$ \\\hline
        SM: IH & $235.9^{\circ} - 296.6^\circ$ & $208.8^\circ - 327.0^\circ$ \\\hline
        MSSM: NH & $299.0^\circ - 15.0^\circ$ & $218.7^\circ - 37.7^\circ$ \\\hline
        MSSM: IH & $234.1^\circ - 308.9^\circ$ & $196.6^\circ - 342.6^\circ$ \\\hline
    \end{tabular}
    \caption{ Viable 1$\sigma$ and 2$\sigma$ HPD intervals for $\delta_{12}^\nu-\delta^e_{12}$ consistent with current experimental values of neutrino observables. RG evolution for the SM and the MSSM are considered for both NH and IH, as discussed in the main text. We note that in all cases the 2$\sigma$ intervals contain $270^\circ = -\frac{\pi}{2}$. }   
    \label{tab:04}
\end{table}

%%%%%%%%%%%%%%%%%%%%%%%%%%%%%%%%%%%%%%%%%%%%%%%
\vspace{0.5cm}
\textbf{Quark phase sum rule:}--
For completeness, let us state that in the quark sector, the analogous calculation (assuming small $\theta^u_{ij}$) yields
\begin{align}
& \delta_{12}^d-\delta^u_{12}
=
\arg\left(  
\frac{s^\textrm{CKM}_{13}-t^\textrm{CKM}_{12}t^\textrm{CKM}_{23}e^{-i\delta^\textrm{CKM}}}{t^\textrm{CKM}_{23}+t^\textrm{CKM}_{12}s^\textrm{CKM}_{13}e^{i\delta^\textrm{CKM}}}
\right).\label{quark-phase-sum-rule}
\end{align}
Expanding in leading order in the small CKM mixing angles reproduced the known result from \cite{Antusch:2009hq} (cf.\ Eqs.~\eqref{quarkPSR0} - \eqref{quarkPSR}). From our MCMC analyses, the results obtained for $\delta_{12}^d-\delta_{12}^u$ are listed in Table~\ref{tab:05}. Interestingly, in both sectors the experimental data points to a difference between the 1-2 phases of about $\pm 90^\circ$, which might hint towards a common origin, as will be discussed in the next section. 

\begin{table}[h]
    \centering 
    \begin{tabular}{|c|c|}\hline
    & $\delta_{12}^d-\delta_{12}^u$ (1$\sigma$) \\\hline
        SM & $84.38^\circ - 91.02^\circ$ \\\hline
        MSSM & $84.38^\circ - 91.05^\circ$ \\\hline
    \end{tabular}
    \caption{The 1$\sigma$ HPD range for $\delta_{12}^d-\delta_{12}^u$    including running effects within the SM and the MSSM scenarios.}
    \label{tab:05}
\end{table}

%%%%%%%%%%%%%%%%%%%%%%%%%%%%%%%%%%%%%%%%%%%%%%%
\section{Unified origin of $\delta^\textrm{PMNS}$ and $\delta^\textrm{CKM}$}\label{sec8}
With the experimental data pointing to a difference between the 1-2 phases $\delta_{12}^\nu-\delta^e_{12}$ and $\delta_{12}^d-\delta^u_{12}$ of about $\pm 90^\circ$, we might think of textures of GUT Yukawa matrices capable of explaining this situation in a simple way. 
In the following, we like to point out a possibility which predicts both CP phases from a single imaginary Yukawa matrix element. We consider an SU(5) GUT, where $Y_d$ and $Y_e$ have the following form:\footnote{We remark that alternatively, one can also choose a texture with $y_{32}=0$ but instead $y_{23}\not= 0$ (and a corresponding term $c_{23}y_{23}$ in $Y_e^T$), leading to the same predictions for the Dirac CP phases.}
\begin{align}
&Y_d=\begin{pmatrix}
0&y_{12}&0\\
y_{21}&i\;y_{22}&0\\
0&y_{32}&y_{33}
\end{pmatrix},\;
Y_e^T=\begin{pmatrix}
0&c_{12}y_{12}&0\\
c_{21}y_{21}&i\;c_{22}y_{22}&0\\
0&c_{32}y_{32}&c_{33}y_{33}
\end{pmatrix},\label{YdYe} 
\end{align}
with all parameters $y_{ij}$ being real (as well as the Clebsch factors $c_{ij}$). Only the 2-2 elements are purely imaginary, all other Yukawa matrix elements are real. We note that in supersymmetric flavour models, with the flavour structure generated via the breaking of family symmetries by ``flavon fields'', such specific values of the CP phases of Yukawa matrix entries can readily be generated with spontaneous CP breaking and the method of ``discrete vacuum alignment'' \cite{Antusch:2011sx}. 
Interestingly, textures of this type can also solve the strong CP problem, as has been pointed out in \cite{Antusch:2013rla}. 

Now, diagonalizing the mass matrices following Eq.~\eqref{Me}, and collecting (12)- and (22)-entries, we get,
\begin{align}
&(M_e)_{12}\approx e^{i(-\delta^e_{12}+\chi^e_2-\chi^{e \prime}_2)}\theta^e_{12}m_\mu + ...\;, \label{eq1}
\\
&(M_e)_{22}\approx e^{i(\chi^e_2-\chi^{e \prime}_2)}m_\mu + ...\;,\label{eq2}
\end{align}
here small mixing angle approximation is adopted (both in the charged-lepton and the down-quark sectors), and we ignore terms proportional to the first generation fermion mass.  Ellipses contain next-to-leading order contributions that are small. Phases $\chi^e_i$ and $\chi^{e \prime}_i$ belong to the left and the right rotation matrices.  Finally, by dividing (12)- and (22)-entries in Eq.~\eqref{YdYe} and comparing with  Eqs.~\eqref{eq1} and \eqref{eq2}, we obtain\footnote{We remark that variations of the above texture could also result in somewhat modified predictions. Ultimately, the full flavour model will have to be analysed in order to obtain the specific precise predictions.} 
\begin{align}
&\delta^{d,e}_{12}=\frac{\pi}{2},
\end{align}
and thus
\begin{align}
&\arg\left(  
\frac{s^\textrm{CKM}_{13}-t^\textrm{CKM}_{12}t^\textrm{CKM}_{23}e^{-i\delta^\textrm{CKM}}}{t^\textrm{CKM}_{23}+e^{i\delta^\textrm{CKM}}t^\textrm{CKM}_{12}s^\textrm{CKM}_{13}}\right)= \frac{\pi}{2},
\\
&\arg\left(  
\frac{s^\textrm{PMNS}_{13}-t^\textrm{PMNS}_{12}t^\textrm{PMNS}_{23}e^{-i\delta^\textrm{PMNS}}}{t^\textrm{PMNS}_{23}+e^{i\delta^\textrm{PMNS}}t^\textrm{PMNS}_{12}s^\textrm{PMNS}_{13}}
\right) =-\frac{\pi}{2},
\end{align}
from Eqs.~\eqref{quark-phase-sum-rule} and \eqref{lepton-phase-sum-rule}, respectively. As discussed above, these relations can be used to calculate the predicted ranges for $\delta^\textrm{CKM}$ and $\delta^\textrm{PMNS}$. The first relation corresponds to $\alpha\approx \pi/2$, compatible with the current experimental data, whereas the second relation  corresponds (using central values of mixing angles and without running) to the predictions $\delta^\textrm{PMNS}\approx 286^\circ$ ($\delta^\textrm{PMNS}\approx 290^\circ$) for $\theta_{23}>45^\circ$ and ($\theta_{23}<45^\circ$). It is remarkable to see that a single imaginary vacuum expectation value can be behind the origin of two CP-violating phases, one in the quark sector and another in the lepton sector. Future more precise measurements of $\delta^\textrm{PMNS}$ (and also of $\delta^\textrm{CKM}$) will tell if this possibility could be realised in nature.

%%%%%%%%%%%%%%%%%%%%%%%%%%%%%%%%%%%%%%%%%%%%%%%
%%%%%%%%%%%%%%%%%%%%%%%%%%%%%%%%%%%%%%%%%%%%%%%
\section{Discussion and conclusions}\label{sec9}
The observation that in the quark sector as well as in the lepton sector, the 1-3 elements of the CKM and PMNS mixing matrices are much smaller than the other elements, motivates the hypothesis that the 1-3 elements of all flavour mixing matrices in the up-quark, down-quark, charged lepton and neutrino sectors
in fact all vanish (approximately) in the flavour basis, i.e.\ that $\theta^{u,d,e,\nu}_{13}=0$ in terms of the most common parametrization. The 1-3 mixing in the observable CKM and PMNS mixing matrices are then induced by the non-commutativity of the individual flavour rotations. 

In this paper, we have revisited mixing sum rule relations in the lepton and quark sectors under the assumption of this ``zero 1-3 flavour mixing hypothesis'', and noted that it provides a minimal assumption under which sum rule relations can be derived that do not depend on unphysical phases.
Assuming that the zero 1-3 flavour mixing hypothesis holds, we have obtained exact versions of the sum rule mixing relations and analysed their implications based on the current experimental data, including effects from RG running. In particular, we have analysed how the existing precise measurement of $\theta_{13}^\mathrm{PMNS}$ allows to derive predictions for $\theta_{23}^\mathrm{PMNS}$ in models with constrained 1-2 mixing  $\theta_{12}^\mathrm{e}$ in the charged lepton sector, from the exact relation between $\theta_{13}^{\mathrm{PMNS}}$, $\theta_{23}^{\mathrm{PMNS}}$ and $\theta_{12}^e$ (cf.\ Eq.~(\ref{lepton-sum-rule-1})). Our numerical results are visualized in Fig.\ \ref{fig:t23}.
As specific examples, we have calculated the predictions for $\theta_{23}^\mathrm{PMNS}$ which arise in classes of Pati-Salam models and SU(5) GUTs that relate $\theta_{12}^\mathrm{e}$ to $\theta_{12}^\mathrm{d}$ (cf.\ Table \ref{tab-predictions for theta23pmns}). Furthermore, under the zero 1-3 flavour mixing hypothesis, we have also considered the exact lepton mixing relation involving $\theta_{12}^{\nu}$ and $\delta_\textrm{PMNS}$ (cf.\ Eq.~(\ref{lepton-sum-rule-2})), and visualized our numerical results, which allow to read off the predictions for $\delta_\textrm{PMNS}$ for given 1-2 mixing $\theta_{12}^{\nu}$ in the neutrino sector, in Figs.\ \ref{fig:delta-withrunning} and \ref{fig:delta-withrunning-MajPhasediff-pi}. We have also discussed the analogous exact relations that hold in the quark sector. 

Let us now look at a few example applications to illustrate the usefulness of our results. For instance, one can immediately judge if some of the the existing GUT flavour models still provide a good fit to the present experimental data on the PMNS parameters. For example, the two SU(5) GUT flavour models presented in \cite{Antusch:2013kna} and \cite{Antusch:2013tta} satisfy the zero 1-3 flavour mixing hypothesis to a good approximation. While \cite{Antusch:2013kna} features a NH, it is inverted in \cite{Antusch:2013tta}. Both models use the Clebsch factor combination ($6,6,\frac{1}{2}$) and thus, following the discussion in section VI and using the Table provided in \cite{Antusch:2018gnu}, predict $\theta_{12}^e \in (14.50^\circ,15.16^\circ)$. From Fig.\ \ref{fig:t23} one can clearly see that the resulting predictions for $\theta_{23}^\mathrm{PMNS}$ are outside the current experimental $3\sigma$ range, so both models - which provided good fits to the data at the time they were constructed - are now disfavoured by the current more precise experimental results. 
From Figs.\ \ref{fig:delta-withrunning} and \ref{fig:delta-withrunning-MajPhasediff-pi} one can also evaluate if models with, e.g. tri-bimaxial or bimaximal neutrino mixing can be viable hypothesis for model building (assuming the charged fermion sectors also satisfy the zero 1-3 flavour mixing hypothesis): one can see that while tri-bimaximal mixing, leading to $\theta_{12}^e \approx 35.3^\circ$, predicts $\delta_\textrm{PMNS}$ inside the current $1\sigma - 2\sigma$ range, models with bimaximal mixing can not be realised in consistency with the data.

Furthermore, we have also derived a novel  ``lepton phase sum rule'' which, however, in contrast to the mixing sum rule relations, only holds (free of unphysical phases) in a small angle approximation for the charged lepton mixing contributions. Confronting both, the quark and lepton phase sum rules with the present data for the PMNS and CKM parameters, we point out that in the context of GUT flavour models, the quark and lepton CP violating phases can both be predicted from a single imaginary element in the mass matrices. 

\bibliographystyle{style}
\bibliography{reference}
\end{document}